# Infrared and NMR Spectroscopic Fingerprints of the Asymmetric $H_7^+O_3$ Complex in Solution

Eve Kozari,[a] Mark Sigalov,[a] Dina Pines,[a] Benjamin P. Fingerhut,[b] and Ehud Pines*[a]

Infrared (IR) absorption in the 1000–3700 cm$^{-1}$ range and $^1$H NMR spectroscopy reveal the existence of an asymmetric protonated water trimer, $H_7^+O_3$, in acetonitrile. The core $H_7^+O_3$ motif persists in larger protonated water clusters in acetonitrile up to at least 8 water molecules. Quantum mechanics/molecular mechanics (QM/MM) molecular dynamics (MD) simulations reveal irreversible proton transport promoted by propagating the asymmetric $H_7^+O_3$ structure in solution. The QM/MM calculations allow for the successful simulation of the measured IR absorption spectra of $H_7^+O_3$ in the OH stretch region, which reaffirms the assignment of the $H_7^+O_3$ spectra to a hybrid-complex structure: a protonated water dimer strongly hydrogen-bonded to a third water molecule with the proton exchanging between the two possible shared-proton Zundel-like centers. The $H_7^+O_3$ structure lends itself to promoting irreversible proton transport in presence of even one additional water molecule. We demonstrate how continuously evolving $H_7^+O_3$ structures may support proton transport within larger water solvates.

## 1. Introduction

Proton solvation in aqueous solutions is paramount for comprehending countless proton transfer reactions and proton transport processes.[1] The four smallest protonated water clusters identified in the gas phase by their distinctive IR absorption spectra are $H_3^+O$, $H_5^+O_2$, $H_7^+O_3$ and $H_9^+O_4$.[2] The solvation patterns of the proton in the liquid phase are fluxional and are difficult to assign to specific geometric structures of the aqueous proton[3] because of the fluctuating solvent environment and active proton transport which continuously affect its aqueous environment.[4] In this study, we probe small protonated water clusters in liquid acetonitrile (ACN),[5] a polar hydrogen-bonding-accepting (HBA) solvent. ACN exhibits ultrafast solvent relaxation times comparable to that of liquid water[6] and mixes with water in all proportions. In ACN/$H_2O$ mixtures the proton resides exclusively in the aqueous portion of the mixtures and the protonated water solvates are stabilized from the outside by the ACN solvent because water is a much better solvent for the proton.[4e–h,5] Importantly, ACN does not protonate to ACNH$^+$ in presence of water as $H_3^+O$ is a much weaker acid than ACNH$^+$: $pK_a$(ACNH$^+$) ≤ −10, $pK_a$ ($H_3^+O$ in ACN) = 2.2.[7]

We introduce protons to ACN/$H_2O$ solutions by very strong mineral acids such as $CF_3SO_3H$, $HClO_4$ and HI which dissociate in ACN/$H_2O$ solutions.:[4e,5,8] the $pK_a$'s of $CF_3SO_3H$, $HClO_4$ and HI are ~2–3 in dry ACN.[4e,5,8,9]

The mineral acids become much stronger acids in presence of even traces of water. Here, we mainly use triflic acid $CF_3SO_3H$ and we use the IR absorption of its S–O stretch vibration to report directly on the acid protonation state (Figure 1).

[a] E. Kozari, Dr. M. Sigalov, Dr. D. Pines, Prof. Dr. E. Pines
Department of Chemistry
Ben-Gurion University of the Negev
P.O. Box 653, Beer-Sheva, 84105, Israel
E-mail: epines@bgu.ac.il
[b] Dr. B. P. Fingerhut
Max-Born-Institut für Nichtlineare Optik und Kurzzeitspektroskopie
Berlin 12489, Germany

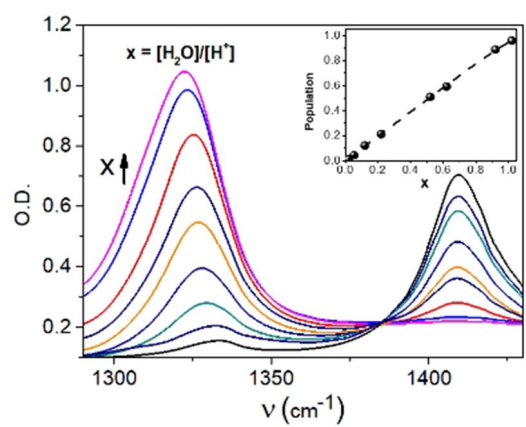

Figure 1. IR absorption of the S–O stretch vibration of 0.6 M triflic acid $CF_3SO_3H$ as a function of the $H_2O$/acid molar ratio x in $CD_3CN$. The peak of the S–OH absorption of the acid is at 1413 cm$^{-1}$ and the peak of the IR absorption of the deprotonated triflate anion $CF_3SO_3^-$ is at 1322 cm$^{-1}$ (from bottom to top: x = 0.02, 0.053 0.12, 0.22, 0.52, 0.62, 0.82, 0.92, 1.02). The acid is fully dissociated at roughly 1:1 acid-to-water ratio: x = 1.02. Inset: A linear correlation is found between x and the fraction of the dissociated acid in the range 0.02 ≤ x ≤ 1. Color coding is for guiding the eye as x varies.

## 2. Results and Discussion

The IR and NMR spectra of 'free' water (Figures 2A, B) considerably change for protonated water as a function of the water-to-acid ratio $x=[H_2O]/[H^+]$ indicating the strong effect of the proton on the $H_2O$ molecules directly solvating it (Figures 3A–C). The spectral changes in our experiments were found to only depend on x and not on the absolute concentrations of water indicating that the observed effects are not due to macroscopic dielectric changes in the solvent. In presence of the proton, the absorption of the OH stretch of the water molecules solvating the proton shifts to the red by about 200 cm$^{-1}$, becomes much wider and loses the distinction between the asymmetric and symmetric OH stretch vibrations. Concomitantly, the 1635 cm$^{-1}$ bend-transition becomes much wider and moves to the blue by about 100 cm$^{-1}$ typical of water molecules solvating the proton in a shared-proton Zundel configuration, $H_5^+O_2$. The 1635 cm$^{-1}$ bend-transition typical of water molecules not strongly interacting with the proton is clearly evident on top of the Zundel absorption for $x \geq 2.2$ but the red-edge of the OH stretch does not resemble the absorption of free water even for $x=3.9$.

The solvent perspective complements the solute perspective of the protonated water clusters shown in Figure 3. The CN triple-bond stretch vibration centered in neat $CD_3CN$ at 2262 cm$^{-1}$ reports on the average proximity of the CN groups to the proton: the closer the electron-rich CN triple-bond is to the proton positive charge the more its absorption blue-shifts.[10] When CN groups solvating protonated water-clusters in ACN are replaced by additional water molecules they move away from the proton and their absorptions red-shift toward their absorption in neat ACN. We thus use the CN stretch as a spectator for the average size of the protonated water clusters which are embedded in ACN. Preparing solutions with larger x values results in an increase in the size of the water solvates and in the absorption of the CN groups less affected by the hydrated proton (Figure 4).

Figure 4 shows four apparent isosbestic points at 2296.5 cm$^{-1}$, 2282 cm$^{-1}$, 2275 cm$^{-1}$ and 2273.5 cm$^{-1}$ that we assign to population transitions between CN groups with IR absorpions centered at 2299 cm$^{-1}$, 2288 cm$^1$, 2278.5 cm$^{-1}$, 2272 cm$^{-1}$ and 2268 cm$^{-1}$. In harmony with the solute perspective of Figure 3 we assign these absorption peaks to CN groups solvating size-distinctive protonated-water solvates. As x increases in the range of $0 \leq x \leq 6$ the absorption of the CN stretch red-shifts and we reconstruct the IR spectra by assuming population transitions between CN groups solvating distinctive protonated water clusters which are affected by x. Importantly, the narrow size-distribution of the protonated water clusters that we find from the analysis of the spectator CN-stretch absorption (Figure 4) and the one we find using the IR absorption of the solute OH stretch, shown in Figure 3, are in full agreement with each other, Figure 5.

Figure 5 clearly indicates that the spectroscopic data emerging from the solvent perspective on the size distribution (cluster-size speciation) of small protonated water solvates in acetonitrile coincides with the size-distribution provided by the solute perspective. Put together, the full agreement between the two independent observations allows us to unequivocally determine the average size of the protonated water solvates in acetonitrile. The cluster-size analysis is further corroborated by NMR spectroscopy done on the same solutions (see below).

We next focus on water/acid solutions in the composition range $2.2 \leq x \leq 2.9$ (Figure 3B). At this solution composition range the average size of the protonated water solvates changes from 2 to 3 water molecules and involves the transition from $H_5^+O_2$ to $H_7^+O_3$ solvates and the solutions are practically made of mixtures of the Zundel cation $H_5^+O_2$, its IR absorption fully characterized in our previous work,[4f–h] and $(H_2O)_3H^+$ solvates which we identify below as the asymmetric $H_7^+O_3$ complex. The cluster-size speciation based on the IR absorption of the clusters was made assuming 50% population of $H_7^+O_3$ at the $x=2.5$ composition with 5% uncertainty. The speciation was further verified independently by analysing the absorption of the CN stretch, Figure 4 and the average chemical shift of the solutions, Figure 6. We reconstruct all chemical shifts in the composition range $2.2 < x \leq 2.8$ assuming it is a weighted average of the chemical shifts of $H_5^+O_2$ and $H_7^+O_3$ their contribution to the total cluster population changing linearly in this range to 80% of $H_7^+O_3$ and 20% of $H_5^+O_2$, see eq. 1 and Figure 6B. Expanding on our NMR experiment we note that protons exchange much faster than the time-scale of the measurement which results in observing a single averaged chemical shift. Under fast-exchange conditions NMR measures a single chemical shift $\delta^1 H_{av}$[11] averaged over all exchangeable H atoms in the protonated water solvates and over the two H atoms of any 'free' water which exchange H-atoms with the protonated water solvates if such water molecules exist in the investigated ACN solutions. Based on the IR and NMR spectra (Figures. 3, 6) we conclude that the solutions are composed of $H_5^+O_2$ and $H_7^+O_3$ solvates with no 'free' water and that their relative abundance changes linearly as a function of x. This gives rise to an average chemical shift which is calculated in Figure 6 B for x=2.4, 2.6, 2.8 solutions according to a two-populations distribution [Eq. (1)]:

$$\delta^1 H_{calc} = (\delta^1 H_{av}(H_5^+O_2)*5*d_2 + \delta^1 H_{av}(H_7^+O_3)*7*d_3)/(5*d_2 + 7*d_3) \quad (1)$$

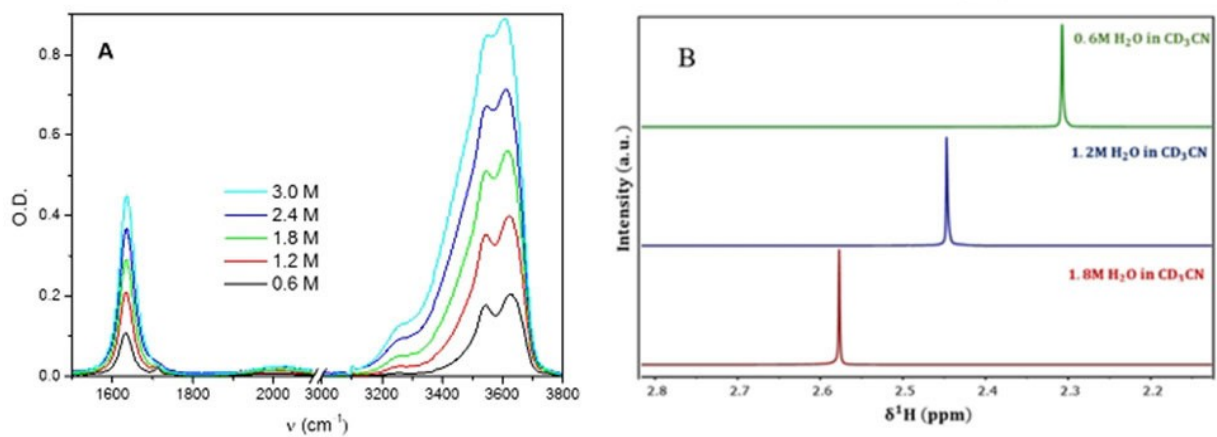

Figure 2. A) The FTIR spectra of unprotonated ('free') water solvated in $CD_3CN$ taken at the indicated concentrations. The bend transition of unprotonated water appears as a relatively narrow absorption at 1635 cm$^{-1}$ and the OH stretch vibration shows a separation between the symmetric and asymmetric vibration in $H_2O$ peaking at 3540 and 3620 cm$^{-1}$ respectively. B) The chemical shifts $\delta^1H^{TMS}$ ($H_2O$) of 0.6 M, 1.2 M and 1.8 M of 'free' water in $CD_3CN$ solutions whose IR absorption is shown in (A). The average chemical shift of the two H atoms in $H_2O$ are at $\delta^1H^{TMS}=2.29$, 2.45 and 2.58 ppm respectively which indicates a small degree of water-water association which slightly increases as a function of the water concentration, $\delta^1H^{TMS}$ of traces of largely unassociated $H_2O$ in ACN=1.92. The corresponding chemical shifts of 0.6 M, 1.2 M and 1.8 M of protonated water protonated with 0.6 M $CF_3SO_3H$ appear at much lower fields[5].

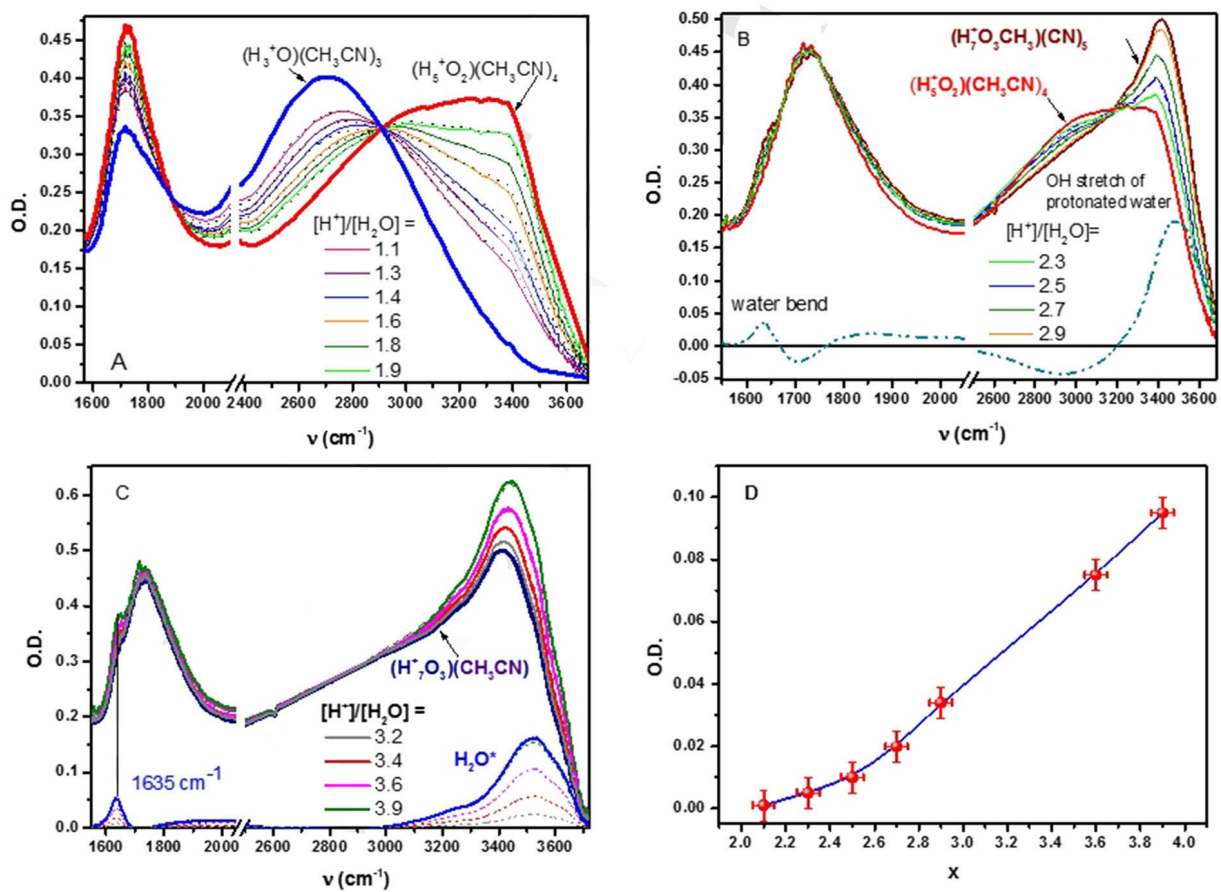

**Figure 3.** The IR absorption spectra for protonated water in acetonitrile. The broken lines are the generated spectral fits of the measured spectra, full lines. The spectral fits are practically indistinguishable from the measured ones. A) Measured IR absorption spectra for solution compositions of $1.1 < x < 1.9$ showing the gradual transition between $H_3^+O$ to $H_5^+O_2$ complexes. The two limiting structures are marked in blue ($H_3^+O$) and red ($H_5^+O_2$). Measured spectra were fitted assuming a linear combination of the two limiting structures with relative populations obeying the stochiometric water-to-acid ratio x and undergoing a two-state transition as a function of x.[4f] B) Top: IR absorption in the OH stretch and the water-bend regions for solution compositions (lower curves to top) $x=2.3, 2.5, 2.7, 2.9$ which is the solution composition range where the transition from $H_5^+O_2$ to $H_7^+O_3$ occurs. Bottom: Difference spectra between the IR absorption of about 90% $H_7^+O_3$ measured in $x=3$ solutions (dark-red) and $H_5^+O_2$ (red-line) highlighting an apparent isosbestic point at 3200 cm$^{-1}$. In the range $2 \leq x \leq 3$ the water bend transition is apparent only above about $x=2.2$. However, our experimental error makes this determination uncertain. C) Absorption spectra for (bottom to top): $x=3.2, 3.4, 3.6, 3.9$ solutions showing no isosbestic point. The core absorption of $H_7^+O_3$ is the most-lower (purple) line and remains invariant upon further increase in the average size of the protonated water solvates. Bottom: The difference spectra between the upper absorptions and the bottom absorption of $H_7^+O_3$ results in a water-like absorption spectrum marked by $H_2O^*$, considerably different from the spectra of 'free' water shown in Figure 2A. The vertical line indicates the water-bend transition at 1635 cm$^{-1}$. D) Analysis of the amplitude of the 1635 cm$^{-1}$ transition as a function of x. We detect the bend-transition for $x > 2.2$ and the correlation appears to approach linearity for $x > 2.6$ extending into water clusters larger than $H_7^+O_3$. Comparing between C and D, the observed O.D. change at the bend-transition between $x=4$ and $x=3$ solutions is about 50% larger than the O.D. change between $x=3$ and $x=2$ solutions. The black curved line is meant to guide the eye, error bars are empiric and are based on the spread in the outcome of multiple sets of experiments.

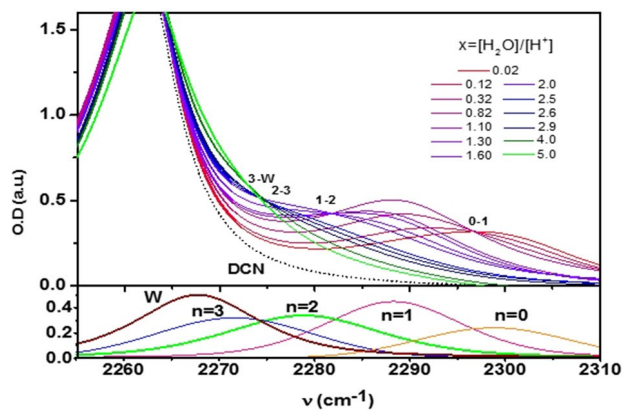

**Figure 4.** The CN triple-bond absorption as a function of the water-to-acid molar-ratio $x=[H_2O]/[H^+]$. The spectral-shifts of the CN triple-bond oscillator are sensitive to the proximity of the CN group to the proton charge and respond by red-shifting to the increase in the size of the protonated water solvates as x increases, see text.[10] n is the order of the water solvate as reported by the CN.

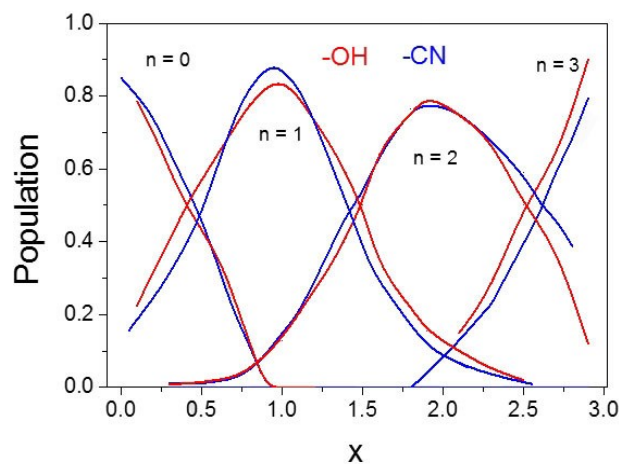

**Figure 5.** A comparison between the solute-side view reported by the OH stretch (red lines) and the solvent-side view reported by the CN stretch (blue lines) of the size-distribution of small protonated water-clusters in acetonitrile as a function of x, the solution composition, in the range $0 \leq x \leq 3$. The analysis of the OH stretch absorption is based on Figure 3A,B and the analysis of the CN stretch is based on Figure 4. The two cluster-size distributions are in full harmony with each other. n indicates the order of the protonated water clusters and the relative normalized abundance of the n=0, 1, 2, 3 clusters are given as a function of x.

We use $\delta^1 H_{av}(H_5^+O_2)=9.5$, and $\delta^1 H_{av}(H_7^+O_5))=8.3$. $d_2$ and $d_3$ are the molar fractions of $H_5^+O_2$ and $H_7^+O_3$ respectively which are directly determined from the value of x. To determine the values of the chemical shifts of the x=2, 2.2 and x=3 solutions we use the full speciation that is derived from our analysis of the OH and CN absorption spectra shown in Figure 5 which also involves averaging the weighted chemical shifts of $H_3^+O$ and the protonated water tetramer $H_9^+O_4$ which were determined separately.[5] We find an almost perfect correlation between the calculated and the measured $\delta^1 H_{av}$ values. The experimental $\delta^1 H_{av}(H_7^+O_3)$ value 8.3 is in excellent agreement with the average theoretical chemical shift of $H_7^+O_3$ in Ref. 5, $\delta^1 H_{av}=8.35$, which is the average $\delta^1 H_{av}(H_7^+O_3)$ value of the symmetric (8.58) and asymmetric (8.12) $H_7^+O_3$ complexes in acetonitrile.[5] Such a good agreement is only possible for a system were all water molecules are protonated and in the form of either $H_5^+O_2$ or $(H_2O)_3H^+$ because the chemical shift of 'free' water (Figure 2B) is too small to be accommodated by Eq. 1. Consequently the theoretical $\delta^1 H_{av}$ calculations were done for pure $H_7^+O_3$ with no 'free' water.[5]

Summarizing our experimental findings, we conclude that: (1) all water molecules in our experiments directly solvate the proton with practically no 'free' water up to at least $x=[H_2O]/[H^+]=4$; (2) the size-distribution of small protonated water solvates in acetonitrile $(H_2O)_nH^+$, n=1,2,3,4 is narrowly distributed around the average number of water molecules per proton, i.e., around x.

### 2.1. Infrared Fingerprint of $H_7^+O_3$ Solvates

Our 3-pronged experimental approach unequivocally demonstrates the existence of small protonated water solvates in acetonitrile. In particular, we are able to determine with high certainty for the first time the IR absorption spectra of the protonated water trimer in these solutions. In Figure 7, the IR absorption spectra between 1500–3750 cm$^{-1}$ are compared for $CF_3SO_3H$ and $HClO_4$ acid using x=3 solutions containing about 90% of all protonated water clusters in the form of $H_7^+O_3$ in $CD_3CN$.

The two practically identical IR absorption spectra consist of a blue-shifted water-bend transition at 1735 cm$^{-1}$ assigned to the two water molecules in $H_5^+O_2$[4] and a very weak much-narrower bend-transition at 1635 cm$^{-1}$ resembling, "free" $H_2O$ in $CD_3CN$ but having a much weaker optical density than the optical density of the "free" water-bend transition measured at the same water concentration. Evidently, the difference in the two acid-anions do not cause a considerable change in the IR spectra of the protonated water solvates. NMR spectroscopy and the CN absorption of the solutions demonstrate that the 1635 cm$^{-1}$ transition shown in Figure 7 belongs to water molecules that are part of the protonated water trimer, otherwise the measured average chemical shifts and the displayed blue shifts in the CN stretch vibration would have been much smaller. In the OH stretch region, the spectrum in $2 \leq x \leq 3$ does not closely resemble either the protonated water dimer, the spectrum of "free" water or any weighted combination of the two,

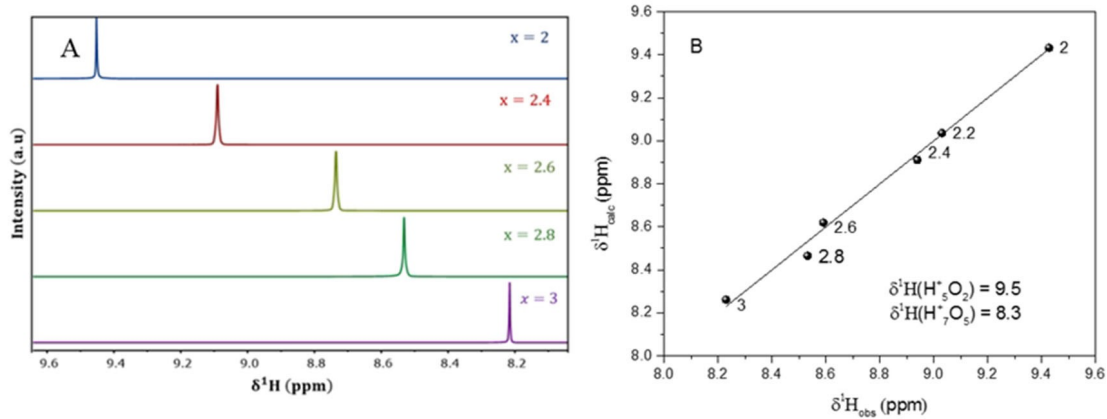

**Figure 6.** A) The measured average chemical shifts $\delta^1 H_{av}$ of 0.6 M $CF_3SO_3H$ in $CD_3CN$ solutions in the range $2.2 \leq x \leq 2.8$ (0.6 M acid in 1.3–1.7 M $H_2O$) at $T = 293$ K, see also Figure 2B to compare with the chemical shifts of unprotonated 'free' water at similar concentrations. B) Comparison between the experimental chemical shifts $\delta 1H_{av}$ and the calculated $\delta^1 H_{calc}$ values (eq. 1) assuming the solution is composed of only $H_5^+O_2$ and $H_7^+O_3$ solvates without any 'free' water. The relative abundance of the two protonated water solvates changes as a function of x and confirms to the population distribution shown in Figure 5.

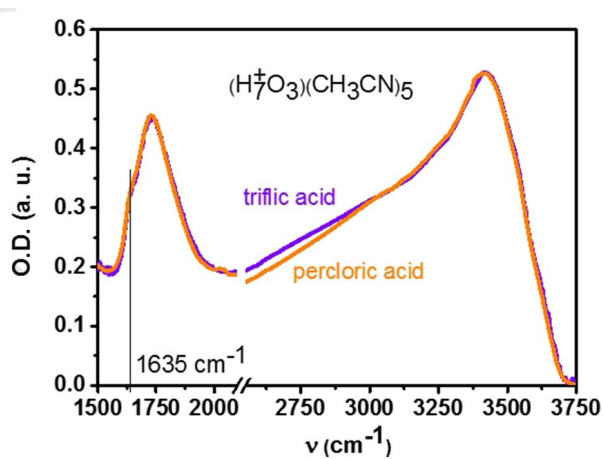

**Figure 7.** IR absorption of largely (90%) $H_7^+O_3$ solvates in $CD_3CN$ using 0.6 M $CF_3SO_3H$ or 0.6 M $HClO_4$ each with 1.8 M $H_2O$ ($x = 3$) showing the OH stretch and water-bend 1500–3750 cm$^{-1}$ regions. The weak 'free-water'-like bend-transition is indicated by the vertical line, however the NMR spectra of the solutions indicate that there is no 'free' water in these solutions.

Figure 3B. We thus assign the spectrum in Figure 7 to a hybrid complex of protonated water, a protonated water solvate which is a combination between the Zundel cation $H_5^+O_2$ and a single, strongly H-bonded water molecule. The strong H-bonding interactions modify both spectra. We identify this hybrid protonated water complex having the $(H_2O)_3H^+$ stoichiometry with the fully solvated $H_7^+O_3$ complex in acetonitrile. Unlike in the gas phase,[2] where the protonated water trimer is symmetric, in solution we expect the very rapidly fluctuating electric field of the solvent to induce an asymmetric distortion of the structure. An asymmetric distortion is apparent in the IR spectra of the complex in the OH absorption region and also comes about in our QM/MM simulations.[4h]

Further support for our structure assignment of the $(H_2O)_3H^+$ solvate to an asymmetric $H_7^+O_3$ complex is given by the IR absorption in the shared-proton region 1150–1450 cm$^{-1}$. Using HI as the proton source allows to directly access this region because the acid anion, I$^-$, does not absorb in this region. We find that in the shared proton region the absorption of $H_7^+O_3$ resembles the shared-proton absorption of $H_5^+O_2$ (Figure 8). The shared proton band slightly blue-shifts from 1145 cm$^{-1}$ ($H_5^+O_2$) to 1160 cm$^{-1}$ ($H_7^+O_3$) and has about 15% larger spectral width at FWHM. The absorption of HI in bulk water further blue-shifts to 1205 cm$^{-1}$ and has a practically identical spectral width at FWHM to that of $H_7^+O_3$ (Figure 8).[4g]

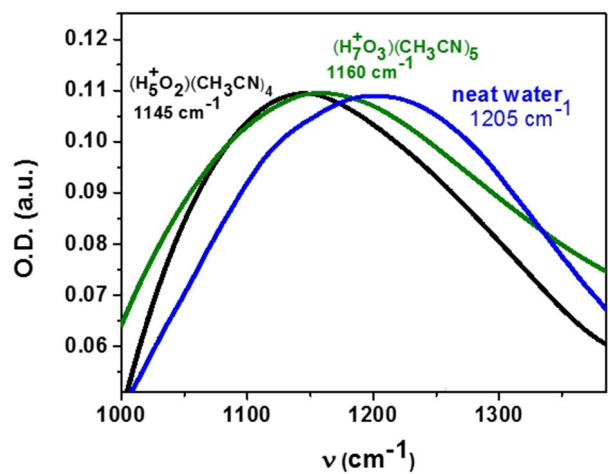

**Figure 8.** IR spectra of 0.6 M HI in ACN in the shared-proton region 1000–1450 cm$^{-1}$ for the indicated protonated water solvates and for 0.6 M HI in pure water. The wavelengths of maximum absorptions are indicated above the curves.

## 2.2. Quantum Mechanical/Molecular Mechanical Molecular Dynamics Simulations

Quantum mechanical/molecular mechanical molecular dynamics (QM/MM-MD) simulations carried out for an extended trajectory[4h] were performed to analyze the structure and electric field induced distortions and proton-translocation dynamics of $H_5^+O_2$, $H_7^+O_3$ and $H_9^+O_4$ solvates in fluctuating environment of ACN solvent at ambient conditions (T = 300 K). The extended long-time trajectory data exceeding 0.5 ns (simulation time of 547.58 ps for $H_9^+O_4$, see Computational Details) allows insight into the $H_7^+O_3$ to $H_7^+O_3$ proton transport dynamics in the $H_9^+O_4$ unit (see below). QM/MM-MD simulations utilize hybrid density functional SOGGA11-X[12] for the QM region. SOGGA11-X/aug-cc-pVTZ level of theory accurately reproduces highest level, wave-function based coupled-cluster benchmark data and previous QM/MM-MD simulations of $H_5^+O_2$ were able to provide infrared absorption spectra of $H_5^+O_2$ in the shared-proton region 1150–1450 cm$^{-1}$ in good agreement with the experiment.[4g]

Our current QM/MM-MD approach is benchmarked via the simulated IR spectra in the OH absorption region of $H_5^+O_2$ and $H_7^+O_3$ (Figure 9) that can be directly compared to our experimental spectra (Figures 3A, B). We find generally good agreement in the peak positions and width of the OH absorption region of $H_5^+O_2$ and $H_7^+O_3$ compared to the experiment (peak positions: 3270 cm$^{-1}$ vs ~3300 cm$^{-1}$ for $H_5^+O_2$ and 3350 cm$^{-1}$ vs ~3375 cm$^{-1}$ for $H_7^+O_3$). Moreover, the QM/MM-MD approach faithfully reproduces the higher intensity of $H_5^+O_2$ in the 2700–3200 cm$^{-1}$ region while for $H_7^+O_3$ the much higher intensity in the 3400–3500 cm$^{-1}$ range is also reproduced. We conclude that the QM/MM-MD method applied on intact $H_5^+O_2$ and $H_7^+O_3$ structures show a satisfactory agreement with the main features of the experimental spectra in the OH absorption region (Figure 3B). The computational model thus provides a reasonable description of both structural and electronic changes accompanying the interaction of $H_5^+O_2$ with the additional water molecule in the $H_7^+O_3$ complex subject to the electric field fluctuations of the ACN environment.

Figure 10 presents the potential of mean force of proton translocation in molecular chains of four water molecules and an excess proton ($H_9^+O_4$). The elementary steps leading to $H_7^+O_3$ to $H_7^+O_3$ proton transport are proton translocation within the $H_7^+O_3$ motif of the $H_9^+O_4$ chain (Figure 10A) and the rearrangement of hydrogen bond network within $H_9^+O_4$ (Figure 10B). As reported before,[4h] QM/MM-MD simulations capture the unique structural properties of the $H_7^+O_3$ motif. The fluctuating solvent field induces a prevailing solvation structure of a dimeric $H_5^+O_2$ unit for the excess proton that gives rise to the characteristic proton transfer mode in the ~1200 cm$^{-1}$ region.[4g,h] The particularly strong hydrogen-bond of the dimeric proton-harboring $H_5^+O_2$ unit within $H_7^+O_3$ has a median O···O distance $R_1 = 2.46$ Å, only about 0.05 Å larger than in the gas phase. The dimeric unit is preserved in the larger water solvates including the star-like structure of $H_9^+O_4$. The shortest hydrogen bond $R_1$ frequently exchanges with one of the closest water neighbors (median O···O distance $R_2 \sim 2.55$ Å) together forming the $H_7^+O_3$ complex. Translocation of the shortest hydrogen bond $R_1$ within $H_7^+O_3$ which switches the location of the dimeric proton proceeds without contraction of the $H_7^+O_3$ motif (Figure 10A). This scenario results in reversible proton translocation within the 3-oxygens arrangement of the $H_7^+O_3$ unit on a typical ~150 fs timescale. A similar timescale has been revealed in time-resolved anisotropy measurements which are sensitive to changes in directionality of the vibrational transition dipole moment.[13]

Figure 10B shows the $H_7^+O_3$ to $H_7^+O_3$ proton translocation in the $H_9^+O_4$ chain-solvate in ACN. The QM/MM-MD simulations demonstrate irreversible proton translocation along a minimal 4 water-molecule arrangement embedded in ACN. Rearrangements of hydrogen bonds in the 3-fold hierarchical hydrogen-bond network around the active proton, marked as hydrogen-bonds $R_1$, $R_2$ and $R_3$ in Figure 10A, B, drive trimer-to-trimer proton transport. The 2-fold hydrogen-bond length-hierarchy of the $H_7^+O_3$ motif in ACN is preserved in the larger $H_9^+O_4$ solvate. The 3$^{rd}$ hydrogen-bond to the 4th water molecule $R_3$ is considerably weaker for both the star- and chain-like arrangements of $H_9^+O_4$ (median O–O distance $R_3 \sim 2.63$ Å). This creates an intrinsic three-fold hierarchy of hydrogen-bonds in the protonated tetramer not observed in gas-phase simulations. The compression of the collective hydrogen bond coordinates $R_2 + R_3$ and correlated displacement along symmetrized proton transfer coordinate $(z_2 + z_3)/2$ characterize the structural changes upon rearrangement of the hydrogen bond network that are required for translocation of the $H_7^+O_3$ motif within the $H_9^+O_4$ chain and thus causing irreversible proton transfer within the chain. Such rearrangement of hydrogen bond network within the four-water molecular chain is rate determining in ACN environment with a free energy barrier of ~421 ± 53 cm$^{-1}$ (1.2 ± 0.15 kcal/mol). Structurally, translocation of the $H_7^+O_3$ is accompanied by an effective O···O contraction along the chain by about 0.3 Å.

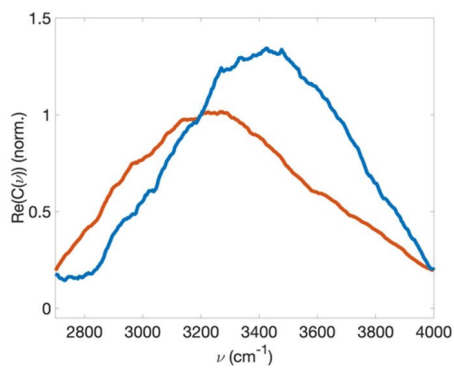

**Figure 9.** Simulated $H_5^+O_2$ (red) and $H_7^+O_3$ (blue) IR spectra obtained from quantum mechanical/molecular mechanical-molecular dynamics (QM/MM-MD) simulations via the Fourier transform $C(\nu)$ of the respective dipole moment autocorrelation function $C(t)$ (see Computational Details). Simulated IR spectra were normalized to the experimental isosbestic point at 3200 cm$^{-1}$ (Figure 3B).

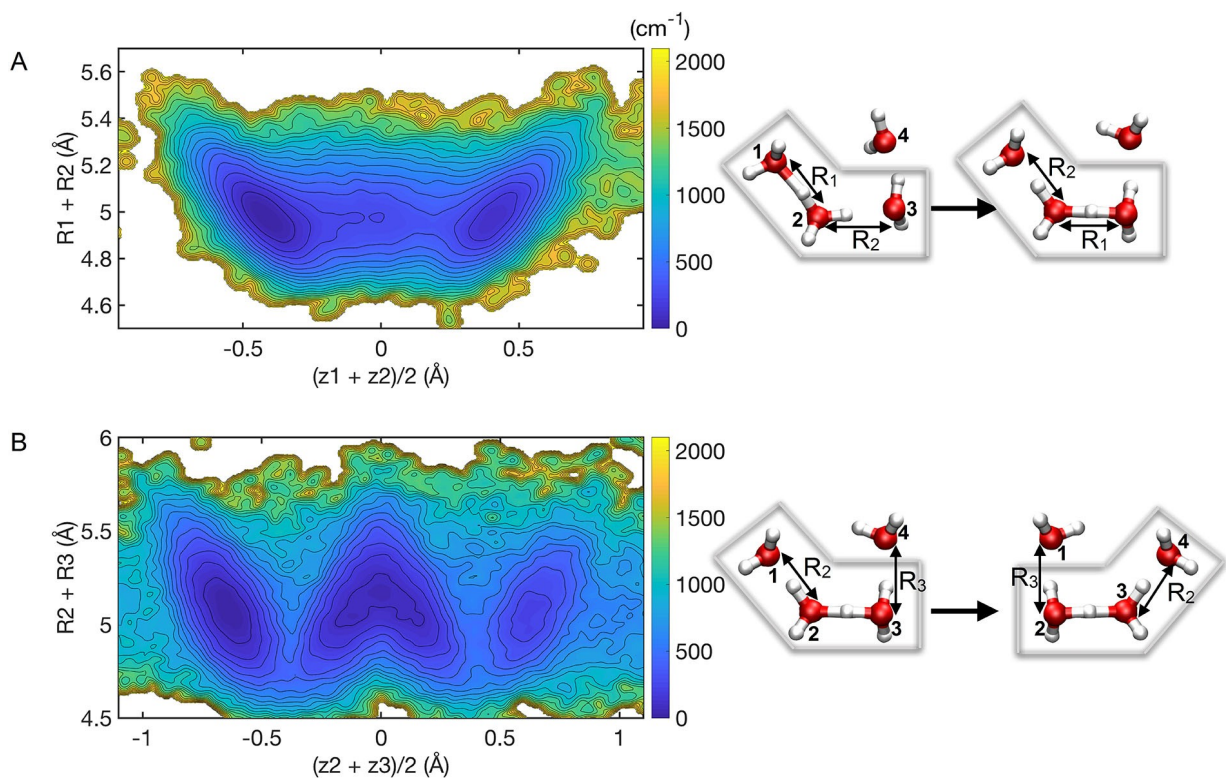

**Figure 10.** Potential of mean force of proton translocation in molecular chains of four water molecules and an excess proton ($H_9^+O_4$): A) Reversible proton translocation within the $H_7^+O_3$ motif is accompanied by fluctuating H-bonding interactions with a 4$^{th}$ water molecule. B) Rearrangement of hydrogen bond network within $H_9^+O_4$ drive irreversible proton transport by translocating the $H_7^+O_3$ unit to a new site with a different geometry. (A) and (B) form the elementary steps of the trimer-to-trimer (T–T) proton transport mechanism. z1, z2 and z3 denote proton transfer coordinates $z = r_{O_i \cdots H^+} - r_{O_j \cdots H^+}$ ($r_{O_i \cdots H^+}$ and $r_{O_j \cdots H^+}$ are distances between $H^+$ and the oxygen atoms of flanking water molecules i and j, respectively) for shortest, second shortest and third shortest O···O distances $R_1$, $R_2$ (A) and $R_2$, $R_3$ (B), respectively. In the limiting geometry of a symmetric dimeric $H_5^+O_2$ motif, the proton resides at equal distances from the two oxygen atoms (z = 0). The symmetrized proton transfer coordinates (z1 + z2)/2 and (z2 + z3)/2 indicate proton translocation within $H_7^+O_3$ and $H_9^+O_4$ motifs, respectively, and the sum of hydrogen bond coordinates $R_1 + R_2$ and $R_2 + R_3$ characterize structural changes. The low correlation of symmetrized proton transfer coordinate (z1 + z2)/2 and the sum of hydrogen bond coordinates $R_1 + R_2$ (A) reflect the minor structural changes of the $H_7^+O_3$ motif upon reversible translocation of the active proton within $H_7^+O_3$ on the ~150 fs timescale.

The QM/MM-MD simulations, supported by our experimental findings of the IR fingerprint of the $H_7^+O_3$ motif, suggest a picture where a fluctuating asymmetric hybrid structure made of a shared-proton unit $H_5^+O_2$ strongly interacting with a 3$^{rd}$ water molecule mediates proton transport via structural diffusion, the so-called Grotthuss mechanism.[1f,14] Persistent translocation of the proton charge,[14,15] i.e., irreversible proton-transport results from translocating the $H_7^+O_3$ unit by one water molecule and is driven by fluctuations in the unit's 1$^{st}$ solvation shell (Figure 11). Such rearrangements shift the center of the trimeric structure from $O_2^*$ to $O_3^*$. The newly formed $H_7^+O_3$ unit (oxygens $O_2$–$O_3^*$–$O_4$) shares 2 water molecules with the $H_7^+O_3$ unit which preceded it (oxygens $O_1$–$O_2^*$–$O_3$).

Adopting a general reaction model by Hynes et al.,[16] water-water hydrogen-bond rearrangement is likely to be clocked by large amplitude angular jumps which results in rotational-like orientational rearrangements of the surrounding water molecules on a typical timescale of 1 ps (pictorially indicated by an arrow in Figure 11).

The translocation of the $H_7^+O_3$ motif is step-wise and allows persistent transport of the proton tightly coupled to ~ps hydrogen-bond angular rearrangements. It should be noted that in this suggested scenario of proton transport, the 3-fold hydrogen-bond hierarchy in the protonated water species is preserved with no transient production or destruction of monomeric $H_3^+O$. Our QM/MM simulations imply that the transition-state for proton transport in aqueous solutions is likely to be associated with the translocation of the $H_7^+O_3$ motif coupled to transient rearrangements of hydrogen-bonds in the trimeric-unit's 1$^{st}$ solvation-shell water molecules in what may be described as streaming $H_7^+O_3$ to $H_7^+O_3$ transitions.

## 3. Conclusions

Our IR and NMR studies have revealed the spectroscopic fingerprints of the core solvation motif, $H_7^+O_3$, of the hydrated proton in ACN/water solutions. QM/MM simulations show that the dimeric shared-proton structure is preserved in fluctuating $H_7^+O_3$ units and suggest it has a central role in structural proton translocation along water chains. The hybrid structure of $H_7^+O_3$ enables-by the merit of its unique hybrid structure-persistent proton transport. It is likely to take part in proton transport through bulk water via structural diffusion, which requires persistent translocation of the proton charge driven by the breaking and formation of hydrogen bonds.[14,15]

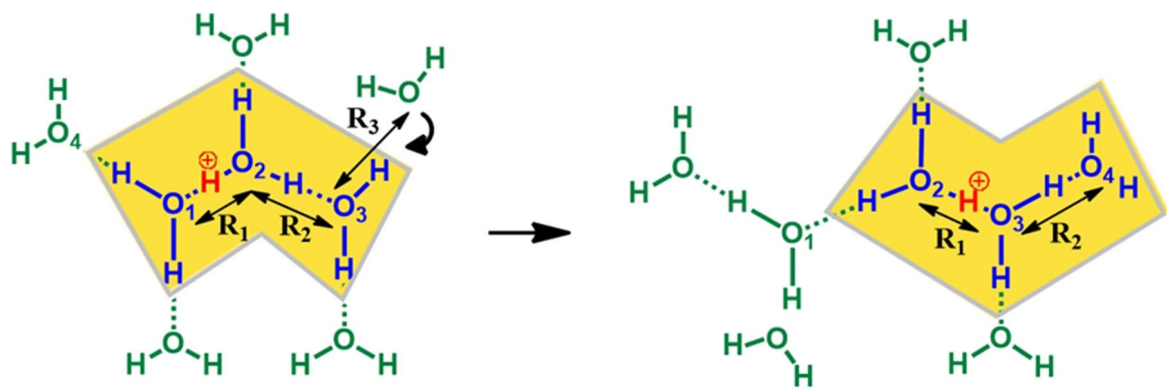

**Figure 11.** A pictorial sketch of the $H_7^+O_3$ complex (blue coloring) in solution allowing for proton translocation by random fluctuations in the H-bonding interactions with its surrounding water molecules drawn for a $[H_2O]/[H^+]=8$ solvate. The trimer-to-trimer proton transport mechanism is highlighted by the orange color. The active proton is indicated in red and its irreversible translocation via the $H_7^+O_3$ unit is driven by random fluctuations in the position of the five water molecules directly solvating the $H_7^+O_3$ unit marked in green. $R_1$ is the dimeric (shortest) O–O separation within the $H_7^+O_3$ complex which migrates by one water molecule as the proton translocate, $R_2$ is the 2nd (intermediate) O–O separation distance within the $H_7^+O_3$ complex and $R_3$ is the longest O–O separation distance between $O_3$ of the $H_7^+O_3$ unit and one of its solvating water molecules approaching $O_3$.

## Methods Section

### Experimental Details

#### Sample preparation

Solution preparation was done in a home-made glob-box under dry Ar atmosphere to minimize the water content in the acetonitrile solution. Anhydrous $CD_3CN$ 99.8% (Sigma-Aldrich), and $CF_3SO_3H$ (triflic acid) 99% extra pure (Strem Chemicals) were used as received. $HClO_4$ (Sigma-Aldrich), was 70.7% by weight ($x=H_2O/acid=2.3$) and was further concentrated down to about $x=1.3$ by drying the concentrated acid in acetonitrile solutions using $P_2O_5$ as the drying agent. HI (Merck) was 67% by weight ($H_2O/acid=3.5$) and the acid was further dried by $P_2O_5$ in acetonitrile solutions down to about $x=1.8$.

#### Infrared spectra measurements

Steady state IR spectra of 0.6 M or 0.5 M of the acids in $CD_3CN$, were carried out with $CaF_2$ windows using optical path lengths of 15 and 25 μm.

The measurements were done with Nicolet IS10 FTIR instrument with 0.4–1 cm$^{-1}$ resolution while applying constant dry $N_2$ flow to keep moisture out of the surrounding atmosphere. The data as shown represent the average of 20 FTIR absorption scans after subtraction of the absorption of the pure $CD_3CN$ solvent.

#### $^1$H NMR experiments

$^1$H NMR experiments were carried out using BrukerBiospin GmbH 400 MHz spectrometer of 0.6 M solutions of triflic acid in deuterated acetonitrile ($CD_3CN$). The measurements were done in a sealed NMR tube, temperature stabilized and cooled by liquid nitrogen. The residual $^1$H impurity signal of $CHD_2CN$ in $CD_3CN$ solvent ($\delta$ $^1$H = 1.94 ppm on the TMS scale) was used as the NMR standard chemical-shift.

### Computational Details

#### Quantum Mechanical/Molecular Mechanical Molecular Dynamics Simulations

QM/MM-MD simulations are employed to simulate the dynamics of protonated water tetramer cluster $H_9^+O_4$ in fluctuating ACN solvent at ambient conditions (T=300 K) and follow the procedure detailed in Refs. [4 g,h]. The QM/MM treatment utilizes hybrid density functional SOGGA11-X (Ref. [12]) that is able to provide infrared absorption spectra of $H_5^+O_2$ in ACN in the shared-proton region in good agreement with the experiment.[4g] The use SOGGA11-X/aug-cc-pVTZ level of theory reproduces highest level, wave-function based coupled-cluster singles-doubles with perturbative triples corrections (CCSD(T)/aug-cc-pVTZ) potential and shows excellent performance for barrier heights, the energetic position of minima for varying $O_1 \cdots O_2$ distances $R_1$ in $H_5^+O_2$ and equally good performance for the variation of $O_2 \cdots O_3$ ($R_2$) in $H_7^+O_3$, i.e., the coordinate of a third water molecule approaching interacting with $H_5^+O_2$.

The QM/MM-MD simulations have been performed with a locally modified version of the AMBER14 program package[17] extending the mixed quantum-classical (QM/MM) module[14][18] to the Molpro program package[19] as external quantum chemistry program. QM calculations of the solvated excess proton with two ($H_5^+O_2$), three ($H_7^+O_3$) and four water molecules ($H_9^+O_4$) have been performed on density functional level of theory with the hybrid GGA functional SOGGA11-X[12] employing density fitting.[20] Water molecules together with the excess proton $H^+$ were entirely treated on QM level. The comparison of the OH absorption region of $H_5^+O_2$ and $H_7^+O_3$ uses trajectory data of Refs. [4 g,h] for simulation of spectra via the Fourier transform C(ν) of the dipole moment autocorrelation function C(t) (Figure 9). Simulated infrared absorption spectra in the OH stretch region were smoothened via 72 frequency-step moving average. Insight into the trimer-to-trimer proton translocation mechanism in $H_9^+O_4$ (Figure 10) is provided via a new extended trajectory data set of 547.58 ps simulation time for $H_9^+O_4$.

The protonated water clusters were placed in a rectangular box of acetonitrile molecules (1180 acetonitrile molecules, box dimension: 45.742×55.084×55.084 Å) employing periodic boundary conditions. Acetonitrile molecules were treated on the molecular mechanics (MM) level with force field parameters of the rigid body united atom model taken from Ref. [21]. The rigid body united atom model reproduces dynamic properties like the solvation correlation function with good accuracy.[22] Triflate ($CF_3SO_3^-$) counter ion (force field parameters and partial charges taken from Ref. [2123]) was employed for charge neutrality. Long-range electrostatic interactions within the MM region were treated with the particle mesh Ewald (PME) approach with real space cut-off (10 Å). Electrostatic embedding in the QM Hamiltonian was performed by including the partial charges of acetonitrile molecules within a cut-off radius $R_c = 10$ Å of the QM region.

Classical equation of motion in QM/MM-MD simulations were propagated with a time step of $\Delta t = 0.25$ fs. The system was initially minimization on the QM/MM level for 3500 steps. Equilibration was performed by heating the system to 100.0 K during a short 10 ps trajectory performed in the nvt ensemble (Langevin dynamics with 1.0 ps$^{-1}$ collision frequency) followed by further equilibration in the npt ensemble at 300 K for 60 ps (1.0 bar reference pressure). During npt equilibration, the density converged to $0.7469 \pm 0.0015$ g/cm$^3$ in reasonable agreement with the experimental value (0.7768 g/cm$^3$). Equilibration was performed with position constraints (5 kcal/mol) on the QM-region to avoid a disintegration of the $H_9^+O_4$ cluster. Starting from the corresponding equilibrated configurations, long nvt QM/MM trajectories were simulated where the position restraints were gradually weakened (5 kcal/mol→0.2 kcal/mol). Production simulations were launched without position restraints every 20 ps from the weakly constrained trajectory. A reduced aug-cc-pVDZ basis set for the QM region was used for equilibration and production simulations without position constraints employed aug-cc-pVTZ basis set (391 contracted basis functions for $H_9^+O_4$). To provide a balanced description of different star like (resembling Eigen-type structures) and wire configurations of $H_9^+O_4$ (E−$H_9^+O_4$ and W−$H_9^+O_4$, respectively) two equilibration trajectories were simulated with position constraints around respective E−$H_9^+O_4$ and W−$H_9^+O_4$ structures. The extended production run trajectory data amount to 547.58 ps simulation time for the water cluster $H_9^+O_4$ in the nvt ensemble at 300 K. Coordinates were saved every 2.5 fs for further investigation. Trimer-to-trimer (T−T) proton transport along a four-water-molecule path was investigated for the protonated tetramer with chain configuration W−$H_9^+O_4$ that accounts for 207.95 ps of the total simulation time (Figure 10B).

Shortest, second shortest and third shortest O⋯O distances $R_1$, $R_2$ and $R_3$, were determined via a 40 time-step moving average (~2 O⋯O oscillation periods) of coordinates $R_{OiOj}$. The procedure follows the character of O⋯O hydrogen bonds but allows for short-time (<1 O⋯O oscillation period) inversion of $R_1$, $R_2$ or $R_3$, thus giving rise to hierarchical and overlapping distributions of O⋯O distances.[4h] We have verified that the results are unchanged for varying moving averages in the range 20–80 time steps.

## Acknowledgements

E.P. acknowledges the support by the Israel Science Foundation grant #1587/16. B.P.F. gratefully acknowledges support through the Deutsche Forschungsgemeinschaft within the Emmy Noether Programme (grant FI 2034/1-1). This project has received funding from the European Research Council (ERC) under the European Union's Horizon 2020 research and innovation programme (grant agreement No 802817).

## Conflict of Interest

The authors declare no conflict of interest.

**Keywords:** hydrated proton · trimer · Zundel cation · CN stretch · molecular dynamics simulations